\documentstyle[aps,epsf]{revtex}

\begin{document}

\title{Suprathreshold stochastic resonance:
suppression of noise by strong periodic signals}

\author{Andrey L. Pankratov}
\address{Dipartimento di Scienze Fisiche "E.R.Caianiello",
Universita' di Salerno, Italy\\ and Institute for Physics of Microstructures
of RAS, Nizhny Novgorod, RUSSIA. \\ E-mail: alp@ipm.sci-nnov.ru.}



\maketitle

\begin{abstract}
We consider an overdamped Brownian motion in "quartic" potential
subjected to periodic driving. This system for the case of
a weak periodic driving has been intensively studied
during past decade within the context of stochastic resonance.
We had demonstrated, that for the case of predominantly
suprathreshold driving the noise in the output signal is strongly
suppressed at certain frequency range: the signal-to-noise ratio
demonstrates resonant behavior as function of frequency.
\end{abstract}

The switching dynamics of an overdamped brownian particle
in a "quartic" potential has been intensively studied in the
past decade in the frame of stochastic resonance (SR)
phenomenon \cite{st}. Stochastic resonance is a nonlinear
noise-mediated cooperative phenomenon wherein the coherent
response to a deterministic signal can be enhanced in the
presence of an optimal amount of noise. It had been observed
in a wide variety of electronic systems, such as, e.g., lasers
\cite{las}, Schmitt triggers \cite{trig}, tunnel diodes \cite{diod}
and SQUIDs \cite{sq1},\cite{sq2}. It is known, that
for a singlelevel threshold systems SR has been observed for
the case of weak (underthreshold) driving \cite{st}
(although residual SR effects are known to occur for
marginally suprathreshold signals \cite{sd}).
In this case as well as in the case of
suprathreshold SR in multilevel threshold systems \cite{NS},
the manifestation of SR is resonant behavior of signal-to-noise
ratio ($SNR$) or other relevant characteristics as function
of noise intensity.

However, most of practical devices are operating in the
predominantly suprathreshold regime, when the transition from one state to
another one over potential barrier occurs deterministically and noise
is the only disturbing factor leading to erroneous switching. As
an example of such a situation we can refer to microwave
hysteretic SQUID \cite{lih},\cite{SQUID}, whose dynamics is
described by the model of Brownian motion in bistable
potential subjected to strong periodic driving at the
given frequency. Such SQUID represents a clear example of
the device, by its very basic idea operating in strongly suprathreshold
regime (its important characteristic is a function of
amplitude of output signal versus the amplitude of input
driving) and for which presence of noise leads to earlier transitions
over potential barrier that results in error of the measured
dc magnetic flux.
For microwave hysteretic SQUID long standing problem is known:
at which parameters it should operate in order to demonstrate
maximal sensitivity. On one hand it is known, that
with increase of pumping frequency the sensitivity of the SQUID
should improve, on the other hand it is known that at frequencies
higher than the cut-off frequency, the performance of the device
should degrade. Some qualitative treatment of this problem has
been done in the past \cite{lih}, \cite{BJ}, but the question is
still opened.

Few papers, where the case of strong periodic
driving was considered both in classical \cite{MS},
\cite{PS} and quantum systems \cite{J}, were addressed to
investigation of area of hysteretic loop and its resonant
behavior as function of frequency of driving signal was
demonstrated. But, to our knowledge, the investigation of
signal-to-noise ratio as function of frequency was not
performed for the case of strongly suprathreshold driving.
This may be explained by the
fact, that the most studies were restricted by adiabatic approximation,
where frequency dependence of $SNR$ could not be investigated
in detail or by linear response theory, where the driving
amplitude was supposed to be small. Nevertheless, in
the frame of linear response theory some weak resonant frequency
dependence of signal-to-noise ratio was recently observed
for a particular case of piecewise rectangular potential \cite{BG}.

Recently we have shown using description via
temporal characteristics, that in a dynamical system
with noise subjected to a strong periodic driving, significant
suppression of noise is possible in a certain frequency range
\cite{PLA}.

The aim of the present paper is to consider
fluctuational dynamics of a Brownian particle in a
"quartic" potential, study the
signal-to-noise ratio and demonstrate that for the
case of a strong (suprathreshold) periodic driving the $SNR$
has resonant behavior as function of frequency.

    Consider a process of Brownian diffusion in a potential profile
\begin{equation}
U(x,t)=bx^4-ax^2+x A\sin(\Omega t+\varphi),
\label{pot}
\end{equation}
where $\varphi$ is initial phase.
\noindent
It is known that the probability density $W(x,t)$ of the Brownian
particle in the overdamped limit (Markov process) satisfies
the Fokker--Planck equation (FPE)
\begin{eqnarray}
{\partial W(x,t)\over\partial t}=
-{\partial G(x,t)\over\partial x}=
{1\over B}\left\{{\partial\over\partial x}\left[
{du(x,t)\over dx}W(x,t)\right]+
{\partial^2W(x,t)\over\partial x^2}\right\}. & &\label{FPE}
\end{eqnarray}
Here $G(x,t)$ is the probability current, $B=\displaystyle{h/{kT}}$,
$h$ is the viscosity (in computer simulations we put $h=1$),
$T$ is the temperature, $k$ is the Boltzmann constant and
$u(x,t)=\displaystyle{U(x,t)/kT}$ is the dimensionless
potential profile. The initial and the boundary
conditions have the following form:
\begin{equation}\label{in}
W(x,0)=\delta(x-x_0), \,G(\pm\infty,t)=0.
\end{equation}

In computer simulations we had chosen the following parameters
of the potential: $b=1$, $a=2$. With such a choice the coordinates
of minima equal $x_{min}=\pm 1$, the barrier height $\Delta U=1$,
the critical amplitude $A_c$ is around $1.5$ and we have chosen
$A=2$ to be far enough from $A_c$. We note, that this is not the
case "just above the threshold level", considered in \cite{sd},
but indeed strong driving: the amplitude $A=2$ is far above the
dynamic threshold. We also performed the analysis for
$A=3; 4; 5$: the results are qualitatively the same, only $SNR$
rises accordingly.

The quantity of our interest is the $SNR$. In accordance with \cite{st}
we denote $SNR$ as:
\begin{equation}\label{snr}
SNR=\frac{1}{S_N(\Omega)} \lim_{\Delta \omega \to 0}
\int\limits_{\Omega-\Delta\omega}^{\Omega+\Delta\omega}
S(\omega) d\omega,
\end{equation}
where
\begin{equation}\label{spec}
S(\omega)=\int\limits_{-\infty}^{+\infty} e^{-i\omega \tau}
K[\tau]d\tau
\end{equation}
is the spectral density, $S_N(\Omega)$ is noisy pedestal at
the driving frequency $\Omega$ and $K[\tau]$ is the correlation
function:
\begin{equation}\label{corf}
K[\tau]=\left< \left< x(t+\tau)x(t)\right> \right>,
\end{equation}
where the inner brackets denote the ensemble average and outer
brackets indicate the average over initial phase $\varphi$.

In order to obtain the correlation function $K[\tau]$ we solved
the Eq. (\ref{FPE}) numerically, using the Crank-Nicholson scheme.

In Fig. 1 the spectral density $S(\omega)$ is presented for $kT=0.1$
(the delta-spikes at 1-st and some higher harmonics are
out of figure in order to enhance the noise part). One can see
that the form of $S(\omega)$ significantly depends on driving
frequency $\Omega$, while the amplitude of the output signal is
monotonically decreasing function of $\Omega$. In order to study
the resonant behavior of spectral density, let us plot the $SNR$
as function of driving frequency $\Omega$. From Fig. 2 one can
see, that $SNR$ as function of $\Omega$ has strongly pronounced
maximum. The location of this maximum at $\Omega=\Omega_{max}$
approximately corresponds to the time scale matching condition:
$\Omega_{max}\approx \pi/\tau_{min}$, where $\tau_{min}$ is the
minimal transition time from one state to another one.
The existence of optimal driving frequency may be explained
in the following way. Let us consider the case of adiabatically
slow driving. If noise is absent, the escape would occur
only after the corresponding potential barrier would disappear.
If we add some small amount of noise, the escape would occur
earlier than in the deterministic case at some nonzero barrier
height. If the driving frequency is increased, the potential
barrier height will decrease faster and the escape will occur
at lower barrier, that is more close to the deterministic case.
In the case, when the driving frequency is higher than the
cut-off frequency of the system, $\Omega\ge \Omega_c$
(where $\Omega_c$ has a dynamical sense: for a given
amplitude of the signal the performance of the system
significantly degrades above certain frequency),
the particle would never escape over potential barrier
in the absence of noise and will remain in the vicinity of
initial potential minimum, since there is not enough time
to reach the basin of attraction of another state.
Therefore, there is some frequency range, where at the given
small noise intensity the escape will occur over the smallest
potential barrier and namely in this case noise has minimal
effect on the system that results to the maximal $SNR$.

In the case when the driving frequency is higher than the cut-off
frequency, $\Omega\ge \Omega_c$, adding some amount of
noise will help the particle to move
to another state and the conventional stochastic resonance may be
observed (see the inset of Fig. 2 for $\Omega=1$).
Let us note, that we were not able to observe such intuitively
obvious thing as increase of $SNR$ for $kT\to 0$ in the case
$\Omega\ge \Omega_c$ (in this case for $kT=0$ we will have
small but nonzero oscillations near the initial potential
minimum) due to divergence of the algorithm for $kT<0.02$
($\Omega=1$) and the last three points, presented in the
inset of Fig. 2, correspond to $kT=0.1; 0.05; 0.02$, respectively.

In conclusion we have shown that in the dynamical system with
noise driven by a strong sinusoidal signal (predominantly
suprathreshold driving) the influence of noise is significantly
reduced in a certain frequency range: the signal-to-noise ratio
is a resonant function of frequency of the driving signal. This
effect is of real importance for applications since it may allow
to operate a concrete device (like a SQUID or other electronic
devices) in the regime of minimal noise-induced error.

For driving frequencies higher than the cut-off frequency
the conventional (but suprathreshold) stochastic resonance
has been observed.

Author wishes to thank Prof. M. Salerno for helpful discussions.
This work has been supported by the MURST (Ministero
dell'Universita' e della Ricerca Scientifica e Tecnologica), by
the INFM (Istituto Nazionale di Fisica della Materia) and
partially supported by the Russian Foundation for Basic Research
(Project N~00-02-16528, Project N~99-02-17544 and
Project N~00-15-96620).

\newpage

\begin{figure}[th]
\centerline{
\epsfxsize=12cm
\epsffile{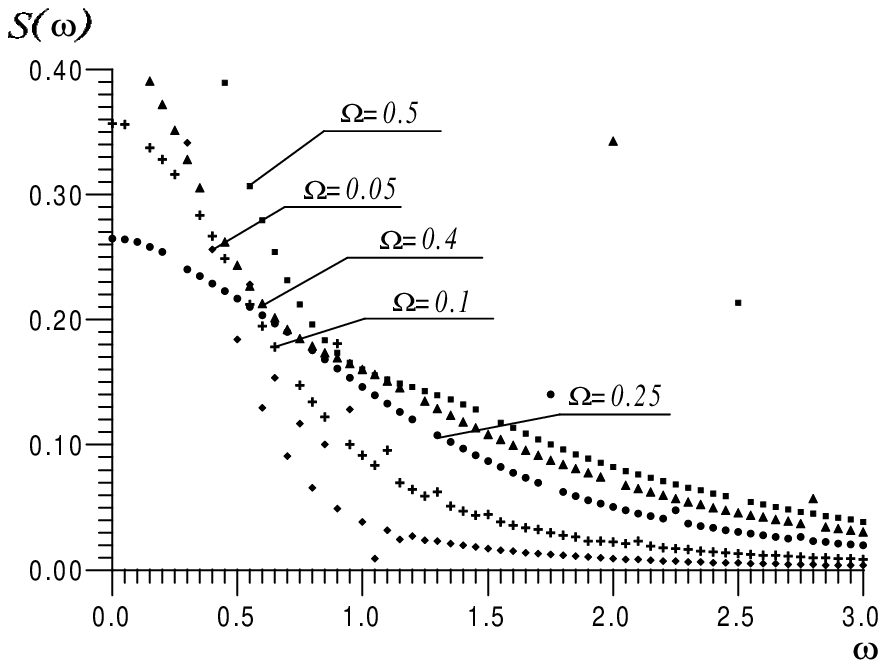}}
\vspace{5pt}
\caption[b]{\label{fig1}
Spectral density $S(\omega)$ with enhanced
noise part, $kT=0.1$, $A=2$.}
\end{figure}

\newpage

\begin{figure}[th]
\centerline{
\epsfxsize=12cm
\epsffile{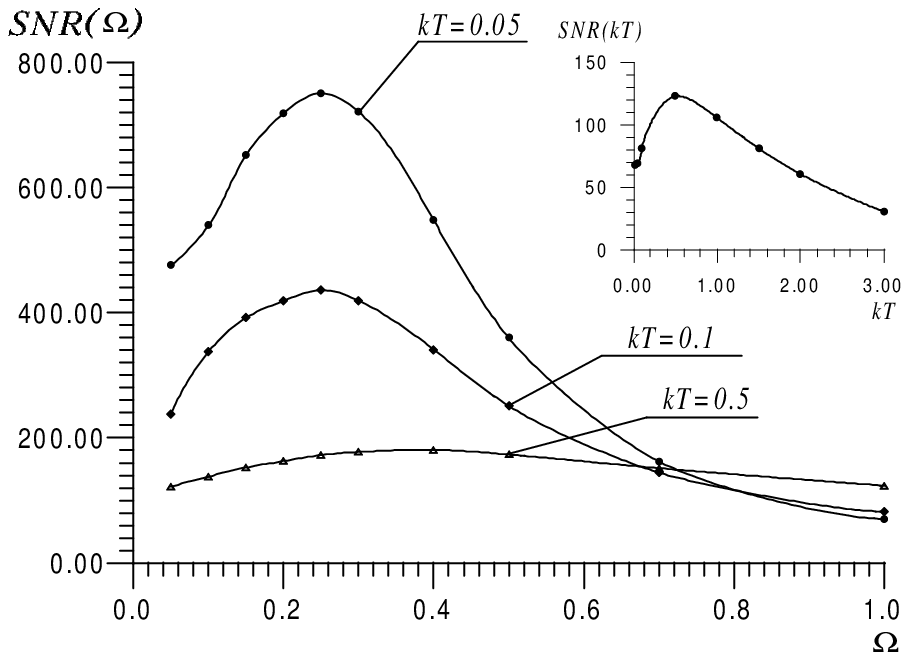}}
\vspace{5pt}
\caption[b]{\label{fig2}
Signal-to-noise ratio as function of driving frequency
for $A=2$. Inset: $SNR$ as function of $kT$ for $\Omega=1$.
}
\end{figure}

\end{document}